\documentclass[aps, pra, showpacs,superscriptaddress , unsortedaddress, twocolumn,
preprintnumbers]{revtex4}
\usepackage{graphicx}
\usepackage{calc}
\usepackage{amsmath}
\usepackage{amssymb}

\newcommand{\D}{\mbox{\rm d}}
\newcommand{\Tr}{\mbox{\rm Tr}}
\renewcommand{\Im}{\mbox{\rm Im}}

\begin{document}
\preprint{PHYSICAL REVIEW A {\bf 80}, 021802(R) (2009)}
\author{A. A. Semenov\footnote{sem@iop.kiev.ua; also at Bogolyubov Institute
for Theoretical Physics, National Academy of Sciences of Ukraine,
and Institute of Physics and Technology, National Technical
University of Ukraine ``KPI''.
 }}

\affiliation{Institute of Physics,
National Academy of Sciences of Ukraine, Prospect Nauky 46, UA-03028
Kiev, Ukraine}

\author{W. Vogel}
\affiliation{Institut f\"ur Physik, Universit\"{a}t Rostock,
Universit\"{a}tsplatz 3, D-18051 Rostock, Germany}

\title{Quantum light in the turbulent atmosphere}

\begin{abstract}
Nonclassical properties of light propagating through the turbulent
atmosphere are studied. We demonstrate by numerical simulation that
the probability distribution of the transmission coefficient, which
characterizes the effects of the atmosphere on the quantum state of
light, can be reconstructed by homodyne detection. Nonclassical
photon-statistics and, more generally, nonclassical
Glauber-Sudarshan functions appear to be more robust against
turbulence for weak light fields rather than for bright ones.
\end{abstract}

\pacs{42.50.Nn, 42.68.Ay, 03.65.Wj, 92.60.Ta}

\maketitle

Nonclassical properties of quantum light have been of great interest
from the viewpoint of fundamentals of quantum physics and for a
variety of applications, such as quantum information processing and
quantum metrology. Special knowledge on  the propagation of quantum
light through the turbulent atmosphere is required in the context of
implementations of quantum cryptography for communication channels
between earth-based stations~\cite{EarthBasedQKD} and between
satellites and Earth-based stations \cite{Villoresi}. The theory is
well established for the propagation of classical light through the
atmosphere, see, e.g.,~\cite{Tatarskii-Fante}, including phenomena
such as beam wander, beam spreading, scintillations, degradation of
spatial coherence, and others. However, nonclassical properties,
such as sub-Poissonian statistics of
photocounts~\cite{NumberSqueez}, quadrature
squeezing~\cite{QuadrSqueez}, non-positivity of the
Glauber-Sudarshan $P$~function~\cite{Nonclassicality, Kiesel}, and
entanglement~\cite{Horodecki} have been little studied in the
context of the propagation of light through the turbulent
atmosphere.

Due to the occurrence of random fluctuations of the refractive
index, the quantum state of light after transmission through the
atmosphere cannot be presented by a single-mode density operator, in
terms of neither monochromatic nor nonmonochromatic modes. The
photocounting statistics of a combination of scattered modes has
been studied by a random modulation of the
intensity~\cite{DiamentPerina,Milonni}. This model has been further
improved~\cite{GammaGamma}. The technique of the photon
wave-function allows one to consider special cases of single-photon
\cite{Paterson} and two-photon \cite{Smith} states. Another
approach, which describes single-photon states, is presented in
Ref.~\cite{Berman}.

In the present contribution we deal with the effects of an
atmospheric transmission channel on any quantum state of light.
General expressions for the quantum state after transmission are
derived. Based on balanced homodyne detection, one may reconstruct
the statistical distribution of the transmission coefficient through
the atmosphere. By repeated reconstruction of the statistical
distribution, one can significantly reduce the atmospheric noise
effects on the quantum state of light.

\begin{figure}[ht!]
\includegraphics[clip=,width=0.95\linewidth]{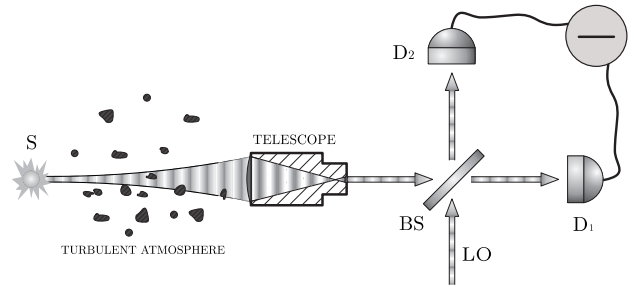}
\caption{\label{Fig1} Homodyne detection of quantum light generated
by the source, S, and propagated through the turbulent atmosphere.
BS is a 50:50 beam-splitter,  D$_1$ and D$_2$ are detectors, LO is
the local oscillator.}
\end{figure}

For dealing with continuous-variable quantum states, we are
considering the experimental setup in Fig.~\ref{Fig1}. The light
from a source is transmitted through the atmosphere and collected by
a telescope (or some other device). Subsequently, a balanced
homodyne detection setup is used to filter out the desired
nonmonochromatic mode from other modes and background radiation by
an appropriate local oscillator, for details
see~\cite{HD,Welsch,SemenovCav}. The remote local oscillator can be
synchronized with the source field by, e.g., the technique of the
optical frequency comb \cite{OFC}. In the limit of a strong local
oscillator, the difference of photocurrents in the detectors is
proportional to the field quadrature of the nonmonochromatic output
mode defined by the local-oscillator pulse. Knowledge of the
quadrature distributions enables one to get the complete information
about the quantum state of the considered output mode. For example,
the method of optical homodyne tomography enables one to reconstruct
the Wigner function, the photon-number distribution, moments of the
radiation field, and the density operator in an arbitrary
representation, for a review see~\cite{Welsch}.

We assume that the propagation of quantum light through the
turbulent atmosphere is a linear attenuating process. The
corresponding losses are caused by light absorption and scattering
as well as by the mode-mismatch due to different shapes of the
output and local-oscillator pulses. The Glauber-Sudarshan
$P$~function~\cite{GlauberSudarshan} of an attenuated
nonmonochromatic output mode, $P_T\left(\alpha\right)$, is related
to the $P$~function of the input mode,
$P_\mathrm{in}\!\left(\alpha\right)$, as (cf., e.g.,
\cite{MandelBook})
\begin{equation}
P_T\left(\alpha\right)=
\frac{1}{|T|^2}P_\mathrm{in}\!\left(\frac{\alpha}{T}\right),\label{IOR_P_1}
\end{equation}
where $T$ is the complex transmission coefficient, with
$\left|T\right|^2\leq 1$. An important difference between turbulent
media and other lossy systems is that the transmission coefficient
$T$ is a random variable with fluctuating phase and magnitude. This
means that the $P$~function of the output mode,
$P_\mathrm{out}\left(\alpha\right)$, is obtained through averaging
$P_T\left(\alpha\right)$ with the probability distribution of the
transmission coefficient (PDTC), $\mathcal{P}\left(T\right)$, as
\begin{equation}
P_\mathrm{out}\left(\alpha\right)=\int\limits_{\left|T\right|^2\leq
1}\D ^2 T\,\mathcal{P}\left(T\right)
\frac{1}{|T|^2}P_\mathrm{in}\!\left(\frac{\alpha}{T}\right).\label{IOR_P}
\end{equation}
The integration is performed over the circular area,
$\left|T\right|^2\leq 1$. This represents the quantum-state
input-output relation for light propagated through the turbulent
atmosphere.

The explicit form of the PDTC should be obtained from a theory,
which considers turbulence properties of the atmosphere as well as
specific conditions of the experiment. Since this is a complex
problem, we may only consider a simple model for the case of small
fluctuations. The corresponding PDTC can be obtained similar to the
probability distribution of the intensity modulation in
Ref.~\cite{DiamentPerina}. For this purpose, we consider a discrete
set of turbulent eddies, each of them is characterized by a random
transmission coefficient $T_k$. The total transmission coefficient
is $T=\prod\limits_k T_k$. The central limit theorem implies that
the PDTC is a two-dimensional distribution, which is log-normal with
respect to the magnitude $t=\left|T\right|$ and normal with respect
to the phase $\varphi=\arg T$, \setlength\arraycolsep{0pt}
\begin{eqnarray}
\mathcal{P}\left(t,\varphi\right)&\approx&\frac{1}
{2\pi t \sigma_\theta\sigma_\varphi\sqrt{1-s^2}}\label{NormalDist}\\
&\times& e^{-\frac{1}{2\left(1-s^2\right)}\left[\left(\frac{\ln
t+\bar\theta}{\sigma_\theta}\right)^2
+\left(\frac{\varphi}{\sigma_\varphi}\right)^2 +2s\frac{\ln
t+\bar\theta}{\sigma_\theta}\frac{\varphi}{\sigma_\varphi}\right]}.\nonumber
\end{eqnarray}
Here, $\bar\theta$ and $\sigma_\theta$ are the mean value and the
variance, respectively, of $\theta=-\ln t$; $\sigma_\varphi$ is the
variance of $\varphi$; $s$ is the correlation coefficient between
$\theta$ and $\varphi$. Without loss of generality we suppose that
the mean value of $\varphi$ is zero. This form of the PDTC can be
used only for $\sigma_\theta\ll \bar\theta$ and $\sigma_\varphi\ll
2\pi$. Contrary to the approach based on the random modulation of
the intensity~\cite{DiamentPerina, GammaGamma, Milonni}, we restrict
the $t$-integration to the range $0\le t \le 1$. Accordingly, the
$\varphi$-integration is restricted to $-\pi \le \varphi  \le \pi $.

Of course, the given model [Eq.~(\ref{NormalDist})] will not
properly describe the turbulence properties of the atmosphere under
general conditions. Due to the lack of a general model, it is
important to
develop a method for the experimental determination of the PDTC. 
Let the input field be prepared in a coherent state
$\left|\gamma\right\rangle$. In this case the PDTC can be expressed
in terms of the characteristic function
$\Phi_\mathrm{out}\left(\beta\right)$, of the $P$~function of the
output state as
\begin{eqnarray}
\mathcal{{P}}\left(T_r,T_i\right)&&=\label{Reconstruction}\\&&\frac{1}{4}\sum
\limits_{n,m=-\infty}^{+\infty}
\Phi_\mathrm{out}\left(\frac{\pi}{2\gamma^\ast}\left[m+in\right]\right)
e^{i\pi\left(mT_i-nT_r\right)},\nonumber
\end{eqnarray}
where $T_r$ and $T_i$ are the real and imaginary parts of the
transmission coefficient, respectively. In an optical homodyning
experiment $\Phi_\mathrm{out}\left(\beta\right)$ can be estimated
from a sample of $N$ photocounting difference events $\Delta n_j$,
cf.,~\cite{Welsch, Kiesel},
\begin{equation}
\Phi_\mathrm{out}\left(\beta\right)=
e^{\frac{\left|\beta\right|^2}{2}}\frac{1}{N}{\displaystyle\sum
\limits_{j=1}^{N}\exp\Big[i\frac{\left|\beta\right|\Delta n_j}{r}
\Big]}.\label{PhiMeasur}
\end{equation}
The amplitude and the phase of the local oscillator are fixed to be $r$ and
$\left(\frac{\pi}{2}-\arg\beta\right)$, respectively.

\begin{figure}[ht!]
\includegraphics[clip=,width=0.95\linewidth]{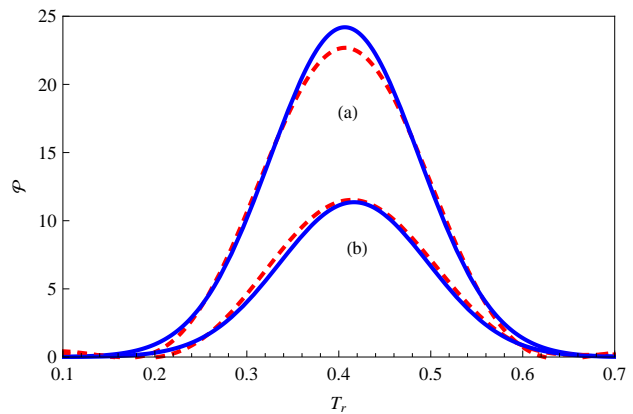}
\caption{\label{Fig3} (Color online) The PDTC is shown for a normal-distribution approximation
of Eq.~(\ref{NormalDist}) with
$\bar{\theta}=0.9 ,\sigma_\theta=0.2, \sigma_\varphi=0.2, s=0.01$,
$T_i=0$ (a), $0.1$ (b). The solid and the dashed lines are the
initially chosen and the reconstructed (from only $5\times 10^3$ sampling
events) PDTC, respectively.}
\end{figure}

To demonstrate the practical usefulness of the reconstruction method
of the PDTC, we have performed the following simulation. We start
with model distribution~(\ref{NormalDist}). For simplicity, it is
approximated here by a normal distribution in the variables $T_r$
and $T_i$. In practice, observed data shall be used, so that this
assumption does not restrict the applicability of our method. We
derive the $P$~function of the output field from Eq.~(\ref{IOR_P})
and calculate the photocount-difference distribution by using the
corresponding integral transformation, see Ref.~\cite{VogelGrabow}.
Now we can simulate the measured data and reconstruct by
Eqs.~(\ref{Reconstruction}) and (\ref{PhiMeasur}) the PDTC. The
result of this procedure is shown in Fig.~\ref{Fig3}, which is in
reasonable agreement with the initially chosen PDTC even for a small
sample of data. In real experiments, this method yields insight into
the true statistics of the turbulent atmosphere.

Alternatively, the PDTC can also be characterized by its statistical
moments. Consider the matrix of normal-ordered moments of the photon
annihilation (creation) operator $\hat{a}$ ($\hat{a}^\dagger$) for
the source field,
\begin{equation}
M_{nm}=\Tr\left [\hat{\rho}\,\hat{a}^{\dagger\,n}\hat{a}^{m}
\right]\equiv \int\limits_{-\infty}^{+\infty}\D^2
\alpha\,P\left(\alpha\right)\alpha^{* n}\alpha^{m}
,\label{MatrixMoments}
\end{equation}
which completely characterizes the quantum state of a light mode,
where $\hat{\rho}$ is the density operator. The moments can be
measured by balanced homodyne detection~\cite{Welsch}. Utilizing the
input-output relation [Eq.~(\ref{IOR_P})], one gets the
corresponding relation between the normal-ordered moments of the
input and output modes,
\begin{equation}
M_{nm}^\textrm{out}= \left\langle T^{\ast n}T^{m}\right\rangle
M_{nm}^\textrm{in},\label{IOR_M}
\end{equation}
where
\begin{equation}
\left\langle T^{\ast n}T^{m}\right\rangle=
\int\limits_{\left|T\right|^2\leq 1}\D^2 T\,
\mathcal{P}\left(T\right)T^{\ast n}T^m\label{MomentsT}
\end{equation}
are the moments of the PDTC.
From Eq.~(\ref{IOR_M}), the moments of the PDTC,
\begin{equation}
\left\langle T^{\ast
n}T^{m}\right\rangle=\frac{M_{nm}^\textrm{out}}{M_{nm}^\textrm{in}},
\label{MomentsMeas}
\end{equation}
are obtained by measuring the moments of the input and output
fields. For example, one can use the input mode in a coherent state
$\left|\gamma\right\rangle$ such that
$M_{nm}^\textrm{in}=\gamma^{\ast n}\gamma^m$. The obtained moments
of the PDTC allow one to determine the nonclassical properties of
the output mode, once the corresponding properties of the input mode
are known.

It is known~\cite{Tatarskii-Fante} that due to the atmospheric winds
narrow light beams are randomly deflected and wide beams scintillate
within a certain time $\tau_\mathrm{atm}$. For data accumulation
times $\tau_\mathrm{data} \gg \tau_\mathrm{atm}$ we expect large
fluctuations of the transmission coefficient $T$. Otherwise, for
$\tau_\mathrm{data} \ll \tau_\mathrm{atm}$ the transmission
coefficient is not significantly fluctuating during the time
$\tau_\mathrm{data}$. However, for this scenario,
$\mathcal{P}\left(T\right)$ is randomly changed between different
series of measurements, separated by time intervals $\Delta
\tau\gtrsim \tau_\mathrm{atm}$. The PDTC and its moments can thus be
permanently monitored by using the method proposed above, which
works with a small sample of data that can be recorded within short
$\tau_\mathrm{data}$ intervals. In this way one can suppress the
influence of long-term atmospheric fluctuations on the quantum state
of the transmitted light.

Let us consider the transmission of sub-Poissonian
light~\cite{NumberSqueez} through the turbulent atmosphere. Using
Eq.~\eqref{IOR_M}, the Mandel parameter (cf.,
e.g.,~\cite{MandelBook}) of the output field, $Q_\mathrm{out}$, can
be related to the Mandel parameter of the input field,
$Q_\mathrm{in}$, as
\begin{equation}
Q_\mathrm{out}=\frac{\left\langle\eta^2\right\rangle}{\left\langle\eta\right\rangle}
Q_\mathrm{in}+\frac{\left\langle\Delta\eta^2\right\rangle}
{\left\langle\eta\right\rangle}M_{11}^\mathrm{in},\label{MandelExpression}
\end{equation}
where $\eta=T^\ast T$ is the efficiency. The first term resembles
the behavior for standard attenuation. The second term is caused by
fluctuations of the efficiency $\eta$ due to the atmospheric
turbulence. It is proportional to the mean photon-number of the
input field, $M_{11}^\mathrm{in}=\left\langle \hat{n}
\right\rangle_\mathrm{in}$. For states of the input field whose mean
photon-number fulfills
\begin{equation}
\left\langle \hat{n} \right\rangle_\mathrm{in}>
-\frac{\left\langle\eta^2\right\rangle}
{\left\langle\Delta\eta^2\right\rangle}Q_\mathrm{in},\label{InequalitySubPoissGen}
\end{equation}
the photocounts of the output mode are always super-Poissonian.
Hence the nonclassical photon statistics of bright quantum light is
destroyed by fluctuations of the magnitude $t$ of the transmission
coefficient, but it is not affected by phase noise.

In other cases, the nonclassical properties of bright light can also
become sensitive to both phase and magnitude noise. In the most
general case a given quantum state is nonclassical if its
$P$~function is not positive definite~\cite{Nonclassicality}. For
weak turbulence, the PDTC has a strong maximum at $T=T_\mathrm{o}$.
Hence the input-output relation [Eq.~\eqref{IOR_P}] reads in the
first-order Laplace approximation as
\begin{equation}
P_\mathrm{out}\left(\alpha\right)\approx\int\limits_{-\infty}^{+\infty}\D
^2\beta\,\frac{1}{|T_\mathrm{o}|^2}P_\mathrm{in}\!
\left(\frac{\beta-\gamma}{T_\mathrm{o}}\right)
\frac{1}{|\gamma|^2}\mathcal{P}\left(\frac{\alpha-\beta}{\gamma}\right)
,\label{IOR_P_Appr}
\end{equation}
where $\gamma=\left\langle \hat{a} \right\rangle_\mathrm{in}$ is the
displacement parameter of the input field and the PDTC is taken in
the Gaussian approximation~\cite{Erdelyi}. If the minimum eigenvalue
of the covariance matrix of the scaled PDTC in
Eq.~\eqref{IOR_P_Appr} obeys $\lambda_\mathrm{min}\geq 2$,
$P_\mathrm{out}\left(\alpha\right)$ represents the Husimi-Kano
$Q$~function~\cite{MandelBook} of the displaced input field combined
with a Gaussian noise, which is always non-negative. Based on the
above approximations, we derive that for any input state with
\begin{equation}
\left|\gamma\right|\geq 2 e^{\bar{\theta}}\sqrt{
\left(\sigma_\theta^2+\sigma_\varphi^2-
\sqrt{\left(\sigma_\theta^2-\sigma_\varphi^2\right)^2+
4s^2\sigma_\theta^2\sigma_\varphi^2}\right)^{-1}}\label{Condition}
\end{equation}
the corresponding output state is classical. For simplicity we have
considered only real $\gamma$ and $\bar{\varphi}=0$, the
generalization to complex $\gamma$ and arbitrary $\bar{\varphi}$ is
straightforward.

\begin{figure}[ht!]
\includegraphics[clip=,width=\linewidth]{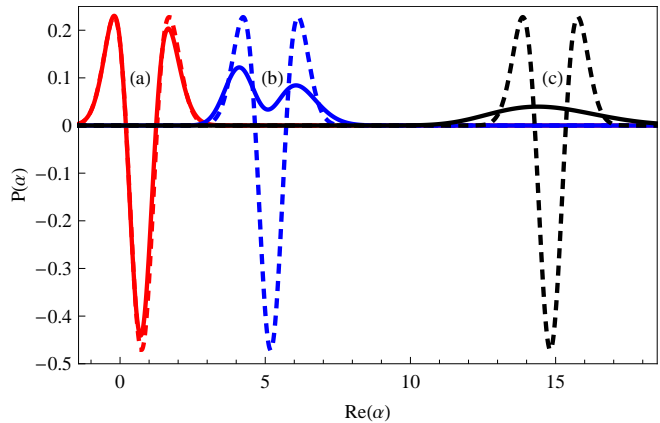}
\caption{\label{Fig2} (Color online) $P$~function of the displaced
SPATS, for $\Im\left(\alpha\right)=0$, $\bar{n}_\textrm{th}=1.11$,
$\gamma=1$ (a), $7$ (b), $20$ (c). The dashed lines show the
standard attenuation for $T=e^{-0.3}\approx 0.7408$. The solid lines
represent the model~(\ref{NormalDist}) for $\bar{\theta}=0.3$,
$\sigma_\theta=0.1$, $\sigma_\varphi=0.14$, and $s=0.01$.}
\end{figure}

As an example we consider the $P$~function of displaced
single-photon-added thermal states (SPATSs)~\cite{Kiesel},
\begin{equation}
P_\mathrm{in}\left(\alpha\right)=\frac{1}{\pi
\bar{n}_\mathrm{th}^3}\left(\left(1+\bar{n}_\mathrm{th}\right)\left|\alpha-\gamma\right|^2-\bar{n}_\mathrm{th}
\right)e^{-\frac{\left|\alpha-\gamma\right|^2}{\bar{n}_\mathrm{th}}},\label{SPATS}
\end{equation}
where $\bar{n}_\mathrm{th}$ and $\gamma$ are the mean number of
thermal photons and the coherent displacement amplitude,
respectively. We compare the $P$~functions of the displaced SPATS
for two cases: (i) the standard attenuation with a fixed
transmission coefficient $T$ in Eq.~\eqref{IOR_P_1}, and (ii)
transmission through the turbulent atmosphere as described by
model~\eqref{NormalDist}. The mean values of the transmission
coefficient are chosen to be equal in both cases. The standard
attenuation does not destroy negativities of the $P$~function. For
small displacements, the situations   in the cases (i) and (ii)
remain similar to each other. However, with increasing displacements
$\gamma$ the atmospheric turbulence destroys the nonclassical
effects, even when they survive for standard attenuation, see
Fig.~\ref{Fig2}.

In conclusion, we have studied the effects of atmospheric turbulence
on the quantum properties of light. It has been shown that the
probability distribution of the transmission coefficient can be
experimentally reconstructed by homodyne measurements. Based on such
a method, one may predict the general turbulence effects on any
quantum state of light. By repeated short-time monitoring of the
PDTC, one can suppress the disturbing effects of long-term
fluctuations on the quantum state of light. Balanced
homodyne detection also allows one to reduce the effects of background
radiation, which is useful for quantum communications under
day-light conditions. We have shown that the
nonclassical effects of bright light fields can be more fragile
against turbulence than for the case of weak fields. A nonclassical
photon-statistics is destroyed by turbulence if the mean photon
number exceeds a critical value. General nonclassicality of the
$P$~function in the output channel is sensitive to the mean coherent
amplitude of the input radiation.

A.A.S. gratefully acknowledges support by NATO Science for Peace and
Security Programme and Fundamental Researches State Fund of Ukraine.
We also thank A.A.~Chumak and J.~Sperling for useful discussions.

\end{document}